# Two-dimensional Cobalt Telluride as Piezo-tribogenerator

*Solomon Demiss[1,2]‡, Raphael Tromer[3]‡, Chinmayee Chowde Gowda[4], Olu Emmanuel Femi[2], Ajit K. Roy[5], Prafull Pandey[6], Douglas S. Galvao[3,7]\*, Partha Kumbhakar[1]\*, Pulickel M. Ajayan[8]\*, Chandra Sekhar Tiwary[1]\**

[1]Department of Metallurgical and Materials Engineering, Indian Institute of Technology Kharagpur, West Bengal, 721302 India.

[2]Materials Science and Engineering, Jimma Institute of Technology, Jimma University, Jimma, Ethiopia.

[3]Applied Physics Department, University of Campinas, Brazil.

[4]School of Nanoscience and Technology, Indian Institute of Technology Kharagpur, West Bengal, 721302 India.

[5]Materials and Manufacturing Directorate, Air Force Research Laboratory, Wright Patterson AFB, OH 45433-7718, United States.

[6]Department of Materials Engineering, Indian Institute of Science, Bangalore 560012, India.

[7]Center for Computational Engineering and Sciences, State University of Campinas, Campinas, SP, 13083-970, Brazil.

[8]Department of Materials Science and NanoEngineering, Rice University, Houston, TX 77005, USA

**ABSTRACT**

Two-dimensional (2D) materials have been shown to be efficient in energy harvesting. Here, we report utilization of waste heat to generate electricity via combined piezoelectric and triboelectric property of 2D Cobalt Telluride ($CoTe_2$). The piezo-triboelectric nanogenerator (PTNG) produced an open-circuit voltage of ~5V under 1N force and the effect of temperature in the 305-363 K range shows a four-fold energy conversion efficiency improvement. The 2D piezo-tribogenerator shows excellent characteristics with a maximum voltage of ~10V, fast response time and high responsivity. Density functional theory was used to gain further insights and validation of the experimental results. Our results could lead to energy harvesting approaches using 2D materials from various thermal sources and dissipated waste heat from electronic devices.

**Introduction**

The rise in global energy demand has led researchers to explore new sources of green energy, which are environmentally friendly and cost-effective. The self-powered systems based on piezo, tribo, thermo, and pyroelectric phenomena are a good source of green energy.[1–4] The energy generated using piezo and triboelectric nanogenerators (PTNG) (consisting of large power densities) is used for small powered appliances such as electronic devices, sensors, and energy harvesting devices.[5–9] As the size of the electronics is reducing, the heat generated during operation is substantial, which can affect the performance of PTNG. Hence, it is essential to study the efficiency of the PTNG at elevated temperature conditions and explore its utilization to harvest energy.[9,10] Recently, considerable innovative research efforts have been dedicated to fabricating piezo/triboelectric nanogenerators using 2D transition metal dichalcogenide (TMDCs) materials.[10–13] Their well-defined structural topologies, good flexibility, and large surface area make them potential candidates for self-powered systems.[13–15] Among TMDCs, Tellurides exhibit unique bandgap sensitivity on the application of strain, which makes them a better candidate for piezoelectric and triboelectric nanogenerators for energy conversion, storage and energy harvesting in small electronic devices.[15–19] Recently, Ajayan et al. has reported the piezoelectric property of 2D Tellurium films (*tellurene*) due to the presence of Born effective charges.[20] Other layered TMDCs such as $MX_2$ (M=Mo, Wo and X=S, Se, and Te) have shown piezoelectricity in their hexagonal form because of their out-of-plane inversion symmetry.[21] Among other layered TMDCs, cobalt-based TMDCs (CoX and $CoX_2$, X=S, Se, Te) have also gained attention due to their catalytic, electrocatalytic, electrical, and magnetic behavior in low dimensions.[22–25] Layered $CoTe_2$ has a hexagonal unit cell, and six Te atoms surround each Co atom. In this work, bulk $CoTe_2$ was obtained using easily scalable and composition-controlled vacuum induction melting. The 2D

CoTe$_2$ was synthesized using a simple sonication-assisted liquid exfoliation method from its bulk crystal. A detailed material characterization using spectroscopy and microscopy has been performed. The energy conversion and energy harvesting property 2D CoTe$_2$ has been carefully investigated by carrying out the experimental and supporting density functional theory (DFT) calculations. The fabricated CoTe$_2$ based PTNG showed excellent response with external strain and temperature, as shown in **Figure 1**. The performance of the PTNG is susceptible to bending, external pressure, and temperature. DFT and *ab initio* molecular dynamics have been employed to analyse the temperature-dependent behaviour of the PTNG. Therefore, the significant output voltage can harvest energy from external stress and waste heat from electronic devices.

**Results and discussions**

**Figure 1** shows the scheme demonstrating simultaneous energy harvesting using a combination of external strain and waste heat dissipated. The PTNG consists of atomically thin CoTe$_2$ exhibiting temperature-dependent behavior. The diffraction patterns of the bulk and exfoliated CoTe$_2$ are as shown in **Figure 2a**. The indexing of these intense diffraction peaks of bulk crystal and 2D CoTe$_2$ were in the orthorhombic crystal structure of the non-centrosymmetric space group *Pnn2* (No. 34) and lattice parameters *a* = 5.32940 Å, *b* = 6.32230 Å, and *c* = 3.90800 Å. XRD pattern indicates that (120) and (140) planes were exfoliated in excess during liquid exfoliation from bulk CoTe$_2$. We observe two polymorphic phases, 1T in bulk form and 1H in the case of 2D, as confirmed from XRD result analysis. Laser light exposure (inset of **Figure 2a**) shows homogenously dispersed of exfoliated CoTe$_2$.

Atomic force microscopy (AFM) measurement has been used to confirm the formation of atomically thin $CoTe_2$ (**Figure 2b**). The thickness was determined from the height profile of the measured histogram plot of the AFM image (**Figure 2c**) and was found to be ~2-4nm. As-prepared bulk and pristine 2D $CoTe_2$ samples were subjected to SEM analysis (see SI, **Figure S1,** and **S2**). The SEM image confirms the exfoliation of bulk $CoTe_2$ into layered $CoTe_2$ with multilayer stacks (**Figure S2**). 2D sheets are further characterized by bright-field transmission electron microscopy (TEM) imaging of $CoTe_2$. Clear sheet-like structures were observed, as shown in **Figure 2d**. The lateral dimension of the sheet was rectangular and ranged from 100–200nm. From the high-resolution TEM image (**Figure 2e-I**) the measured lattice spacing of exfoliated sample was 0.27 nm corresponding to (120) plane. The FFT pattern confirms the orthorhombic structure of $CoTe_2$, as shown in **Figure 2e-II**. The elemental mapping of the phases was done using EDX, confirming the presence of Co and Te. EDX analysis (**Figure S3**) shows only the presence of Co and Te with no impurities. Raman spectroscopy was carried out to confirm the formation of ultrathin $CoTe_2$. **Figure 2f** shows the Raman spectrum of 2D $CoTe_2$ with characteristic vibration modes $A_1$, $B_1$, and $E_g^2$ modes observed. The inset shows multi modes (two $E_g$ and one $A_1$) in the Raman spectrum. The existence of these peaks confirms the formation of a pure phase of $CoTe_2$.[26]

**Figure 2g** shows the X-ray photoelectron spectroscopy (XPS) survey spectra of 2D $CoTe_2$ with the characteristic peaks of Co, Te, O, and C in the samples. De-convoluted XPS peaks of the sample revealed for Co 2p (**Figure 2h**) at 780eV and 796eV corresponding to $CoTe_2$ and other broader peaks were due to surface oxide formation with Co during exfoliation. Peaks in the region of Te 3d (**Figure 2i**) were at 576 and 587eV corresponding to $Te^{4+}$ state (Te $3d_{5/2}$ and Te $3d_{3/2}$). The results are in good agreement with previous studies for the same $CoTe_2$ composite structure.[27,28]

In **Figure 3a-b**, we present the optimized structure of the 2D CoTe$_2$ unit cell, which contains 6 atoms (4Te+2Co). The corresponding lattice parameters were obtained with SIESTA (and Quantum Espresso) code corresponding to an orthorhombic unit cell with lattice vectors given by 3.03 and 8.07Å along X and Y, respectively. We observed five different bond length values: 2.45 and 2.66Å for Co-Te, 2.98 and 3.17Å for Te-Te and, 2.90Å for Co-Co. Thus, there are two different bond types for Co-Te and Te-Te. In **Figure 3b-c**, we present the 2D CoTe$_2$ electronic band structure and the corresponding projected density of states (PDOS) for the optimized structure shown in **Figure 3a-b**. 2D CoTe$_2$ presents metallic characteristics where the weight of 3$d$ atomic orbital is higher than others. We also investigated the effects of Co and Te vacancies and the system preserves its metallic behavior. Our simulations included spin polarization effects, and we verified that the system has an antiferromagnetic characteristic because the energy difference between the ferro and anti-ferro configurations being positive,

$$\Delta E = E_{ferromagnetic} - E_{anti-ferromagnetic} = 0.4 \text{eV} \qquad (1)$$

The fabrication of PTNG using atomically thin CoTe$_2$ has been schematically illustrated in **Figure 1**. **Figure 4a** shows the open-circuit voltage as a function of periodic force ($F$) of 1N. The maximum output voltage reaches up to ~5V. Also, we have calculated the response time of the fabricated cell, and it was found to be ~4ms (**Figure 4b**). The generated output voltage from PTNG by the repetitive finger (index and middle finger) press and release motion is as shown in **Figures 4c** and **4d**. We have also compared the results to bulk CoTe$_2$ samples, and a ten-fold increment of output voltage under the periodic force of 1N (**Figure S4**) has been observed. By increasing the applied pressure, (different finger), the 2D CoTe$_2$ nanogenerators produced an output voltage of ~0.56V and increased up to ~1.6V, as shown in **Figure 4d**. The fabricated nanogenerators have also shown great flexibility (**Figure 4e**) and produced output voltage for bending measurements.

A linear increase in output voltage was observed which was directly proportional to the bending angle (**Figure 4e**). Additionally, a maximum voltage of ~1.5V for the 2D CoTe$_2$ based cell has been achieved when the cell was connected over a bridge rectifier to obtain a positive voltage cycle (**Figure S5**). The output voltage of PTNG was also recorded as a function of external load resistance, as shown in **Figure S6**. It is observed that PTNG has a maximum internal resistance (R$_L$) of ~9MΩ. The maximum output power density (P$_{max}$), output electrical energy (E$_{elec}$), total input mechanical energy (E$_{mech}$), and the energy-conversion efficiency ($\eta$) were calculated to know the efficiency of the fabricated PTNG[29] and the detailed calculations are given in the supplementary section. P$_{max}$ of ~2.56 mWm$^{-2}$ was obtained for the PTNG at the R$_L$ of 9MΩ. The E$_{elec}$ was calculated to be ~4.3 x 10$^{-8}$J at a maximum output voltage of ~5V. Whereas E$_{mech}$ for an applied force of 1N was 5.24 x 10$^{-7}$J. Therefore, the maximum conversion efficiency of the PTNG was found to be ~8.2 %.

In order to investigate the elastic properties of the 2D CoTe$_2$, we replicated the unit cell (**Figure 3a**) of by 2x2x1 along with the X and Y directions, building one orthorhombic supercell composed of 36 atoms. In **Figure 4f,** we present the stress-strain curve for 2D CoTe$_2$. We observed a significant anisotropy for the X and Y directions (see also **Figure 4g**). The Young's modulus was calculated considering the linear regime for strain value until 1%. The estimated Young's modulus values (from the linear regime of the stress/strain curve) were 163.73 and 72.13GPa for the X and Y directions, respectively. We also analyzed the charge distortion effects in the elastic regime to obtain the piezoelectric coefficients using the expression:

$d_{ijl} = P_i/\sigma_{jl}$, (2)

where $d_{ijl}$ is the piezoelectric coefficient for charge variation along the *i* direction, for stress applied along the *jl* direction, and $P_i$ is the polarization vector along the *i* direction. We can use a compact notation for piezoelectric coefficient given by $d_{ij}$, where *i* =1−X, 2−Y, 3−Z and, j=1−XX, 2−YY, 3−ZZ, 4−XY, 5−XZ, 6−ZY.

The charge generation mechanism of the output voltage of a PTNG under external stress is shown in **Figure S7**, which is based on the piezoelectricity generated by 2D CoTe$_2$ and triboelectricity produced from the contact electrification process between Kapton and electrodes. Schematically we have presented the working principle of the as-fabricated PTNG. In order to explain the power generation mechanism for the exfoliated CoTe$_2$ based PTNG, we used a similar theoretical model as reported by Yousry *et al.*[30] **Figure S7 (I, II)** depicts the piezoelectric model, when mechanical stress is applied on a polarized piezoelectric material with non-centrosymmetric nature, the centers of the positive and negative charges shifted, as confirmed from DFT calculation (see SI for more details). The charges generally move from one electrode to another until a point is reached where the potential difference between the electrodes is in a balanced (equilibrium) condition, as in **Figure S7 (I, II)**. The piezoelectric effect decreases when the external force is released, and the electrons flow back to the top and neutralize the positive potential. The triboelectric effect produces substantial electrical output when the two surfaces are not in contact. The pressing action leads to stretching in the device, leading to a piezoelectric polarization. Simultaneously, as a result of this, polarization free charges flow through the external circuit, maintaining the balance in the potential. Appropriate connections of the external electric circuit can generate an enhanced and reversible output signal with electrons moving in the same direction. All the experimental results are due to the synergetic contribution of the triboelectric and piezoelectric effect, and the results are well interpreted.

Telluride-based structures are widely researched for their thermoelectric behavior. Therefore, 2D CoTe$_2$ can also be used as an energy harvesting device using temperature as the source. The utilization of waste heat dissipated for the generation of energy can account for waste heat management. **Figure 1** schematically depicts the measurement of temperature-dependent behavior of 2D CoTe$_2$. The temperature of the fabricated device was monitored through a laser Thermometer. Meanwhile, the device output voltage at each changing temperature was recorded using digital oscilloscope. With the slightest increase in temperature (range of 304K up to 363K), the output voltage linearly increased (**Figure 5a**). After removal of heat source, the maximum voltage attained is reduced to a minimum, gradually. **Figure 5b** represents the average values of the output voltage at increasing and decreasing temperatures. It was observed that the output voltage increased almost linearly and is directly proportional to the temperature change (**Figure 5c**). The variation of polarization of surface charges on the application of heat tends to generate significant thermionic vibrations; these vibrations are responsible for producing a drastic variation in voltage. These produced charges are, in turn, responsible for the temperature-dependent piezoelectric effect that is observed in 2D CoTe$_2$. An open-circuit voltage of ~10V has been obtained from PTNG, under a temperature variation from 305K up to 363K. We calculated the variation of output electrical energy for increased temperature values at constant input mechanical energy of ~5.24 x 10$^{-7}$J. It was observed that for the temperature increase from 305K till 320K there was a gradual increase inefficiency (**Figure S8**). After that, the rate of efficiency increasing was tempered down. The CoTe$_2$ based PTNG device shows a four-fold increment in conversion efficiency at maximum temperature compared to room temperature (RT). Theoretical studies were also performed in the same temperature region to analyze the origin of the higher energy conversion efficiency of the PTNG cell. Therefore, this temperature range was feasible for low-grade-temperature device operation in non-ambient conditions indicating internal overheating issues.

Electronic devices used in our day-to-day activities radiate temperatures in the range of ~300K to 325K on constant usage. One such example is a Laptop; we demonstrate that energy can be harvested from overheating of these devices. The average operating temperature of a laptop is below 313K and above this temperature, it informs internal problems of the device. In order to harness this energy, the fabricated PTNG was attached under a laptop (keyboard and touchpad), where most of the heat flow (overheating) is obtained during the working condition. To evaluate performance of the fabricated PTNG cell, the generated output voltage was measured directly during the working condition of the laptop, as depicted in **Figure 5d**. The possibility to harvest waste-heat energy dissipated from the laptop by installing the described system on the source device was demonstrated successfully. The PTNG shows increased output voltage as a function of laptop temperature. A linear change is observed in the generated relative output voltage ($\Delta V/V_0$), with respect to increasing temperature (**Figure 5e**). It indicates the generation of voltage in the PTNG cell, which is due to temperature dependent piezoelectric effect under the given experimental conditions. The responsivity ($R_S$) of a device is also a key parameter, and it was calculated using, $R_S = \frac{(\Delta V/V_0)}{T}$, where $V_0$ is the initial output voltage at RT, $\Delta V$ is voltage change at temperature ($T$) (**Figure 5e**). It is keenly noted that the device was responsive to temperature variations and exhibited a positive linear relationship in the temperature range of RT up to 321K with a response of 2.013K$^{-1}$. Therefore, the results confirm the application of our PTNG cell for scavenging waste heat dissipated during overheating of a laptop.

Bulk materials generally harness large amounts of heat and temperatures; usually small amounts of heat dissipated go unnoticed. Further, the fabricated device was also sensitive as a function of laptop weight despite being a temperature sensing unit. Here, we have taken a different laptop and monitored their output voltage as a function of weight and heat generation (**Figure 5f**). The calculated responsivity as a function of applied load ($F=mg$, $m$=weight of the laptop) also shows a linear behavior. Therefore, the fabricated lightweight and highly-sensitive PTNG cell can be used for monitoring the temperature of electronic gadgets.

The coupling of piezoelectric and triboelectric properties of $CoTe_2$ creates a polarized electric field with charge separations observed in few layers of 2D $CoTe_2$, which is a result of the time-dependent change in temperature.[9,31] To explore the possible origin of this output voltage, we have used theoretical calculation on the application of strain and temperature. In **Figure 5g-h**, we present the 3D electron density plot (at left) from a top view and electron density map (at right) from a side view. For a small strain of 1% within the linear regime and temperature of 320K, we observe the charge distortion responsible for the changes in the electrical dipole value. In **Table 1** we present the piezoelectric coefficients considering the linear elastic regime for a strain value of 1%. The results were obtained using optimized geometries from DFT calculations at T = 0K and *ab initio* molecular dynamics for finite temperatures. We observed the same trend for piezoelectric coefficient discussed in the present manuscript only for stress along the x-direction, as can be seen in **Table 1** for $d_{11}$, $d_{21}$ and, $d_{31}$. These cases occur as an increase of the coefficient values when the temperature increases. With these above coefficients, we were further able to calculate the current density for the varied temperature conditions (details in supplementary section).[32] Therefore, the current density ($I_{sc}$) was calculated, and a maximum current density of ~51 $nAm^{-2}$ for 300K and

~67 nAm$^{-2}$ for 320K has been obtained. The above theoretical simulation is consistent with experimental observation.

**Conclusions**

In conclusion, we could demonstrate a high-performance, flexible piezo-triboelectric nanogenerator using 2D CoTe$_2$ as an active material. This device was used as a mechanical energy harvesting unit, and it also acts as a self-powered nanogenerator. 2D CoTe$_2$ samples were prepared by liquid exfoliation method, and its piezo-triboelectric property was systematically studied. The experimental results demonstrate that the piezo and triboelectric performance of the exfoliated CoTe$_2$ shows an exemplary output voltage. Additionally, by fruitfully utilizing the thermal property of CoTe$_2$, a four-fold enhancement in energy conversion efficiency has been achieved at maximum temperature. We further analyzed the voltage generation mechanism of fabricated PTNG using DFT calculation. Also, the temperature-dependent behavior on the output performance of the PTNG cell was investigated. To explore the possible origin of this output voltage we have used theoretical calculation in combination with application of strain and temperature. An effective output voltage was obtained using the weight and heat dissipated from a laptop as an energy source. However, we believe that this new type of 2D CoTe$_2$ based nanogenerator has great potential for energy harvesting/conversion applications from thermal sources.

## ASSOCIATED CONTENT

**Supporting Information:** The Supporting Information is available free of charge Experimental section, SEM and EDX of Bulk $CoTe_2$, SEM image of exfoliated $CoTe_2$, EDX of 2D $CoTe_2$, the output voltage of the fabricated cells, Hybrid cell output across bridge rectifier circuit, the output voltage of the PTNG across the different external load resistance, Proposed schematic representation, Energy conversion efficiency as a function temperature, Detailed calculation of power and energy conversion efficiency of PTNG cell.

## AUTHOR INFORMATION

**Corresponding Author**

*Email addresses  ajayan@rice.edu, chandra.tiwary@metal.iitkgp.ac.in, parthakumbhakar2@gmail.com, and galvao@ifi.unicamp.br

**Author Contributions**

The manuscript was written through contributions of all authors. All authors have given approval to the final version of the manuscript. ‡These authors contributed equally.

**Conflict of Interest:** The authors declare no competing financial interest

## ACKNOWLEDGMENT

SD thanks Jimma Institute Technology, Jimma University and Ministry of Science and Higher Education Ethiopia for their funding support. We also thank Indian Institute of Technology Kharagpur for laboratory and characterization facility support. This work was financed in part by the Coordenacão de Aperfeiçoa- mento de Pessoal de Nível Superior - Brasil (CAPES) - Finance Code 001, CNPq, Brazil, and FAPESP, Brazil. RT, CFW, and DSG thank the Center for Computational Engineering & Sciences (CCES), Brazil at Unicamp for financial support through the FAPESP/CEPID Grant 2013/08293-7.

**Figures**

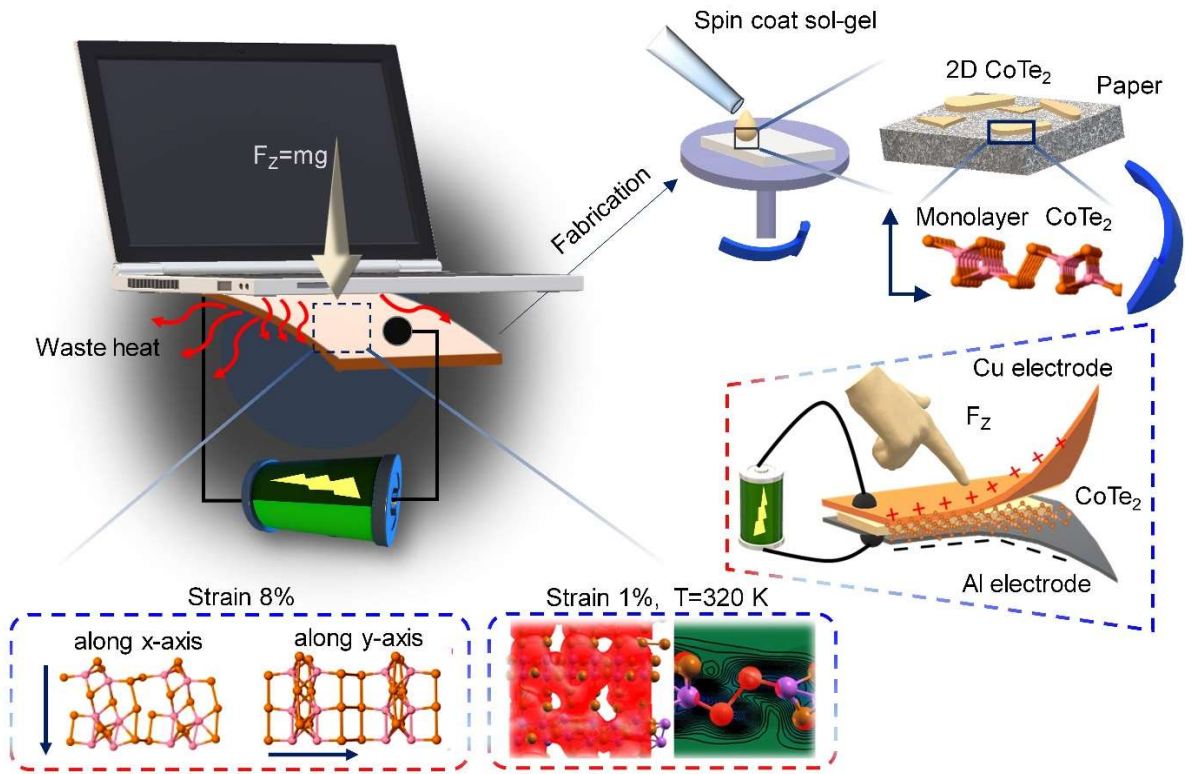

**Figure 1.** Schematic representation of the utilization of waste heat by generating electricity and energy harvesting concept of piezo-tribo cell using 2D CoTe$_2$.

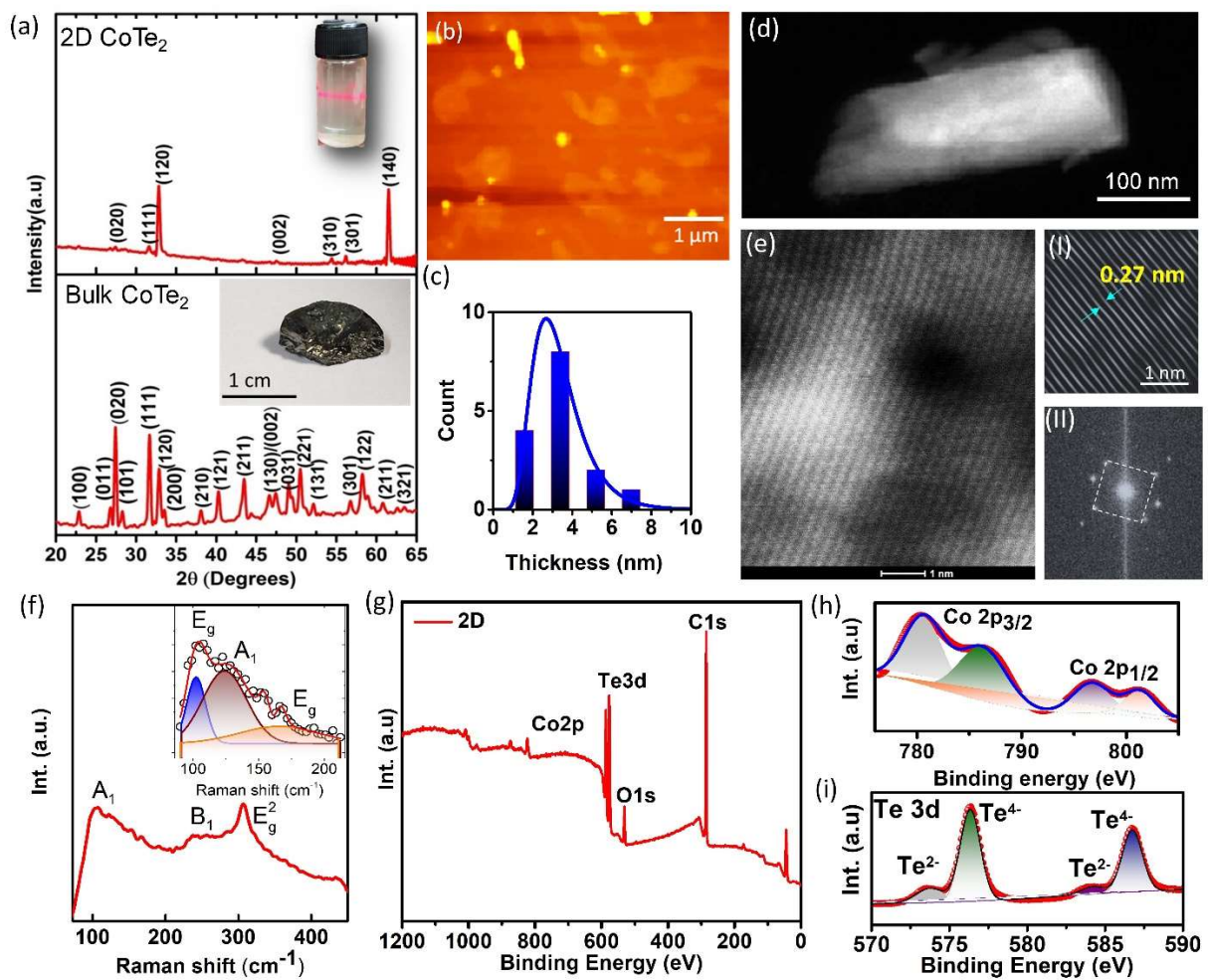

**Figure 2.** (a) The XRD pattern of bulk and 2D CoTe$_2$. The digital image shows bulk CoTe$_2$ (inset-lower) and dispersion of exfoliated CoTe$_2$ in IPA solvent (inset upper), (b) AFM image of 2D CoTe$_2$, (c) Histogram plot as a function of thickness, (d) Bright-field TEM image depicting 2D CoTe$_2$, (e) HRTEM image of the exfoliated sample, (I) The upper inset shows the lattice fringe of 2D CoTe$_2$, (II) The lower inset shows the FFT pattern, (f) The Raman spectrum of 2D CoTe$_2$ (g) Full scan XPS spectra of exfoliated CoTe$_2$, (h) De-convoluted XPS spectra of Co 2p and, (i) Te 3d peak.

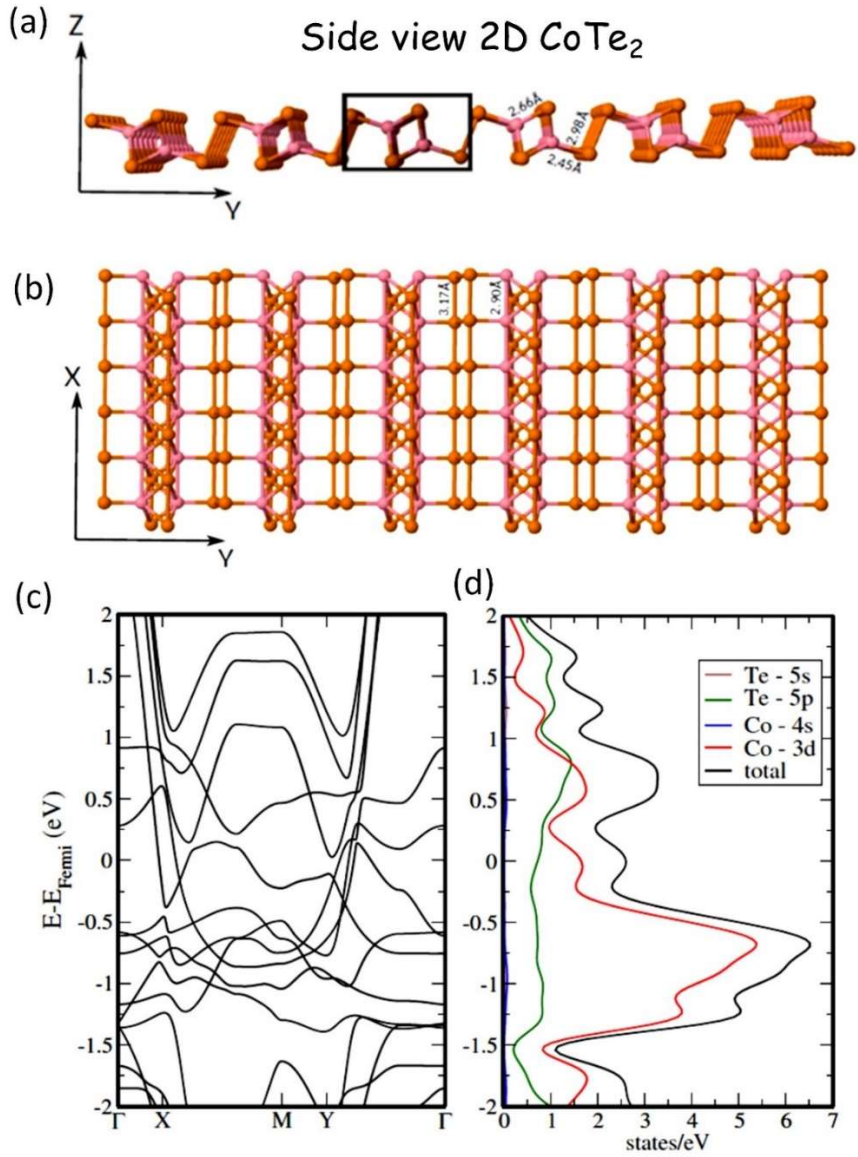

**Figure 3.** (a) Side view and (b) Front views of the optimized structure of 2D $CoTe_2$ exfoliated at the [120] crystal direction from $CoTe_2$ bulk. (c) Electronic band structure, and (d) the corresponding projected density of states (PDOS) of the structure.

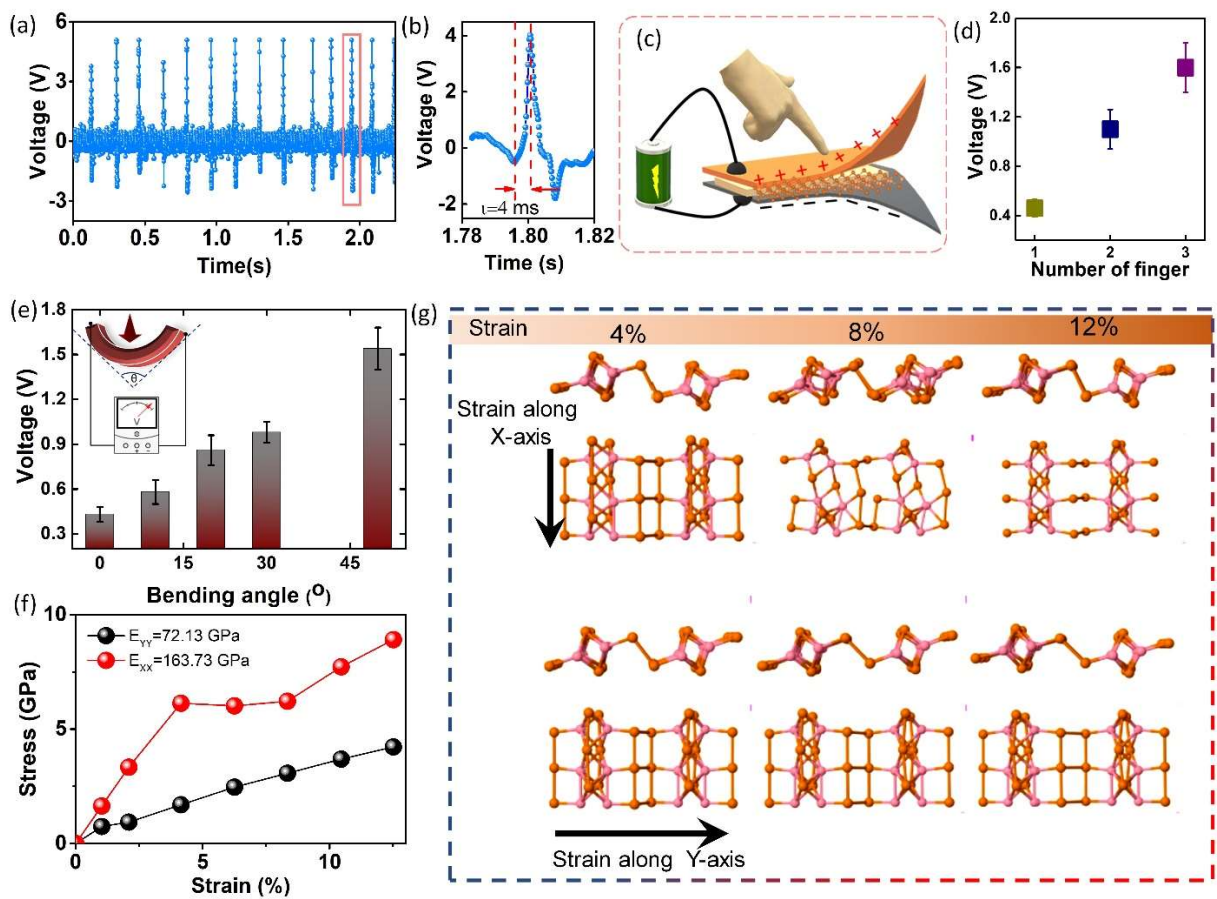

**Figure 4**. (a) The output voltage of the as-fabricated cell, (b) enlarged view of a single pulse, (c) Schematic diagram of the voltage generation using finger press as a pressure source, (d) Output of the open-circuit voltage as a function of applied pressure for different finger pressures, (e) Output voltage response for different bending angles. The inset shows the schematic diagram of the bending measurement connected with a voltmeter, and (g) Stress-strain curve for 2D $CoTe_2$ considering one supercell composed of 36 atoms. The Young's modulus values for each direction are 163.73 and 72.13 GPa along the X and Y directions, respectively.

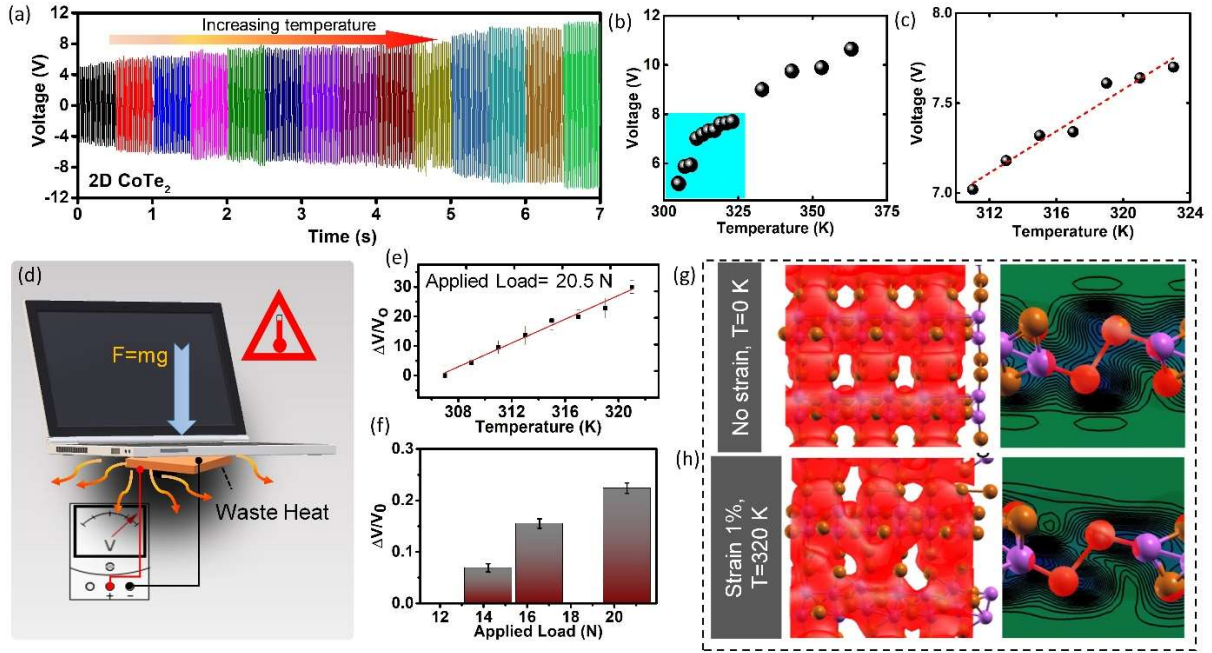

**Figure 5.** (a) Open-circuit voltage response as a function of temperature, (b) Variation of the average output voltage of the cell at different temperatures (increasing and decreasing) under a constant load, (c) Linear fitted curve of output voltage vs. Temperature, (d) The schematic illustration for the temperature measurement and sensing the temperature of a laptop, (e) Relative voltage changes of $CoTe_2$ cell under various temperatures, with a sensitivity of 2.013 K$^{-1}$ (room temperature up to 321 K), (f) Relative voltage changes of the cell as a function of the applied load. The electron density plot at left from the top view and an isosurface value of 0.16e/Å$^3$ and an electron density map at right from a side view at (g) 0 K, no strain, and (h) 320K, 1% strain.

**Table 1**: Piezoelectric coefficients at the linear elastic regime for strain applied along the x, y, and xy-directions, respectively.

| Coefficients (pC/N) | T=0 K | T=300 K | T=320 K |
| --- | --- | --- | --- |
| $d_{11}$ | 5.93 | 11.27 | 13.42 |
| $d_{21}$ | 0.53 | 4.03 | 5.68 |
| $d_{31}$ | 0.08 | 1.6 | 1.97 |
| $d_{12}$ | 26.14 | 20.11 | 21.30 |
| $d_{22}$ | 14.60 | 7.78 | 6.92 |
| $d_{32}$ | 3.93 | 2.76 | 2.97 |
| $d_{14}$ | 5.11 | 13.83 | 0.6 |
| $d_{24}$ | 0.65 | 11.39 | 5.90 |
| $d_{34}$ | 0.30 | 6.55 | 0.84 |